\documentclass[10pt,twocolumn, conference]{IEEEtran}

\usepackage{amsmath}
\usepackage{amssymb}
\usepackage[dvips]{graphicx}
\usepackage{setspace}
\usepackage{epsfig}
\usepackage{amsmath}
\usepackage{amssymb}
\usepackage[dvips]{graphicx}
\usepackage{epsfig}
\usepackage{color}

\newcommand{\tSNR}{\text{SNR}}

\newtheorem{theo}{Theorem}

\newcommand{\figsize}{0.48}

\newcommand{\sr}{\text{SR}}

\begin{document}
\IEEEoverridecommandlockouts
\title{Effective Capacity in Cognitive Radio Broadcast Channels}
\author{\authorblockN{Marwan Hammouda, Sami Akin, and J\"{u}rgen Peissig}
\authorblockA{Institute of Communications Technology\\
Leibniz Universit\"{a}t Hannover\\
Hannover, Germany 30167\\
Email: \{marwan.hammouda, sami.akin, and peissig\}@ikt.uni-hannover.de}\thanks{This work was partially supported by the European Research Council under Starting Grant--306644.}}	
\date{}

\maketitle

\begin{abstract}
In this paper, we investigate effective capacity by modeling a cognitive radio broadcast channel with one secondary transmitter (ST) and two secondary receivers (SRs) under quality-of-service constraints and interference power limitations. We initially describe three different cooperative channel sensing strategies with different hard-decision combining algorithms at the ST, namely OR, Majority, and AND rules. Since the channel sensing occurs with possible errors, we consider a combined interference power constraint by which the transmission power of the secondary users (SUs) is bounded when the channel is sensed as both busy and idle. Furthermore, regarding the channel sensing decision and its correctness, there exist possibly four different transmission scenarios. We provide the instantaneous ergodic capacities of the channel between the ST and each SR in all of these scenarios. Granting that transmission outage arises when the instantaneous transmission rate is greater than the instantaneous ergodic capacity, we establish two different transmission rate policies for the SUs when the channel is sensed as idle. One of these policies features a greedy approach disregarding a possible transmission outage, and the other favors a precautious manner to prevent this outage. Subsequently, we determine the effective capacity region of this channel model, and we attain the power allocation policies that maximize this region. Finally, we present the numerical results. We first show  the superiority of Majority rule when the channel sensing results are good. Then, we illustrate that a greedy transmission rate approach is more beneficial for the SUs under strict interference power constraints, whereas sending with lower rates will be more advantageous under loose interference constraints. Finally, we note that the methodology and the approach we consider in this study can be easily applied into a more general cognitive radio broadcast channel model with more than two SRs.
\end{abstract}

\section{Introduction}
Due to ever-increasing demand for wireless spectrum practices, the concept of cognitive radios emerged as a means to provide transmission solutions by furnishing the idea of secondary users (SUs) in the system. Since then, different complex cognitive radio scenarios have been investigated from several research perspectives. For instance, cooperative strategies for cognitive radio networks attracted significant attention \cite{Latif}. Because of the possible non-continuous presence of the primary (legal) users (PUs) in the environment, the SUs have to sense the transmission channel. Therefore, both non-cooperative and cooperative channel sensing strategies have become the focus of some of the cognitive radio research. Through cooperative channel sensing methods, multiple SUs share their channel observation data with each other in order to boost the channel sensing performance. In that perspective, in an earlier study \cite{Cabric2006}, an experimental research comparing cooperative channel sensing with different channel sensing methods was conducted. Besides, it was shown that the probability of missing the available channels can be made arbitrarily small in independent and identically distributed (i.i.d.) fading channels by increasing the number of the cooperating SUs, while at the same time protecting the PUs in the environment from the harmful interference induced by these SUs \cite{Amir2007}. Furthermore, several decision combining algorithms for the cooperating SUs were studied in \cite{Quan2008} and \cite{Peh2010} as well.

Similarly, cognitive radio broadcast channels took a serious consideration and were explored by many researchers profoundly. For instance, a cognitive radio broadcast scenario in which one primary transmitter and one secondary transmitter are communicating with their respective receivers was considered, and the corresponding achievable regions were presented \cite{ozan}. Further cognitive radio broadcast channel studies were also conducted in \cite{l_li} and \cite{Xiao} where the ergodic sum rate capacity of the SUs with multiple antennas at both the secondary transmitter and the secondary receivers was derived. Finally, Asghari \emph{et al.} studied the adaptive time and power allocation policies by maximizing the achievable capacities of cognitive radio broadcast fading channels \cite{asghari}. In addition to above studies, QoS regarding the buffer and delay constraints has been considered as a vital metric in cognitive radio research as well. Necessarily, effective capacity \cite{Wu2003}, which provides the maximum arrival rate that a service process can support while satisfying the desired QoS constraints, was investigated under average and peak power constraints in cognitive radio relay channels \cite{musavian}, and with imperfect channel sensing results under interference power limitations \cite{Sami2010}.

In this paper, we investigate the throughput of the SUs in cognitive radio broadcast settings where the SUs initially detect the activities of the PUs cooperatively, and then one secondary transmitter performs data transmission to two secondary receivers under QoS and interference power constraints. Especially, unlike in \cite{Deli2013} in which the effective capacity of a broadcast channel with only one single transmitter and many receivers was considered, we address the effective capacity of a cognitive radio broadcast channel where the SUs are engaged in data transmission under the channel uncertainty caused by the channel sensing errors, and the interference power constraints dictated by the PUs. It is worth to mention that our approach can be easily generalized into the cognitive radio broadcast channel models with more than two receivers.

\section{Channel Model}\label{System_Model}
\begin{figure}
\begin{center}
\includegraphics[width=0.3\textwidth]{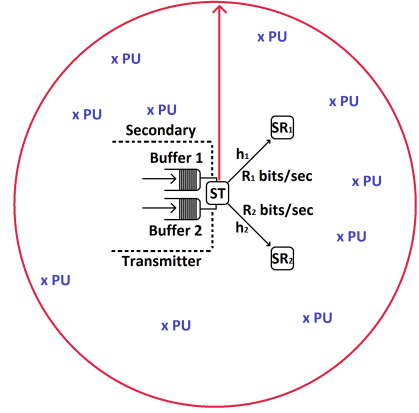}
\caption{Cognitive radio broadcast channel model.}\label{system_model}
\end{center}
\end{figure}

As seen in Figure \ref{system_model}, we assume that a secondary transmitter, denoted by ST, performs communications with two secondary receivers (SRs), i.e., $\sr_{1}$ and $\sr_{2}$, in an environment where there are PUs that are likely to be active occasionally. Therefore, the SUs initially implement channel sensing in order the detect the activities of the PUs. Then, depending on their channel sensing results, they select their data transmission power and rate policies. At the beginning, the data sequences generated by one source (or sources) to be conveyed to the SRs are stored in two different data buffers before the data transmission is performed in frames of $T$ seconds. During the data transmission, the input-output relation between ST and $\sr_{j}$ at the $k^{th}$ time instant is given by
\begin{equation}\label{input_output_idle}
y_j(k)=h_j(k)x(k)+n_j(k)\quad k=1,2,\cdots,
\end{equation} 
when the PUs are inactive, and it is given by
\begin{equation}
\label{input_output_busy}
y_j(k) = h_j(k)x(k) + n_j(k) + s_j(k)\quad k=1,2,\cdots,
\end{equation} 
when the PUs are active. Note that $j$ is a subscript indicating the number of the SRs, i.e., $j\in\{1,2\}$. In (\ref{input_output_idle}) and (\ref{input_output_busy}), $x$ and $y_j$ are the complex channel input at ST and the complex channel output at $\sr_{j}$, respectively. We remark that $x$ carries information to both SRs. Besides, $\{n_j\}$ is a sequence of additive thermal random noise samples at $\sr_{j}$, which is zero-mean, circularly symmetric, complex Gaussian distributed with variance $\mathbb{E} \{|n_j|^2\} = \sigma_{n,j}^2$. Meanwhile, $h_j$ represents the fading coefficient between ST and $\sr_{j}$, which is likewise assumed to be a zero-mean, circularly symmetric, complex Gaussian distributed random variable with variance $\mathbb{E} \{|h_j|^2\} = \mathbb{E} \{z_j\} = \sigma_{h,j}^2$. Note that $z_j$ is the magnitude square of the instantaneous fading coefficient $h_{j}$, and that $h_{1}$ and $h_{2}$ are independent of each other. Furthermore, $s_j$ in (\ref{input_output_busy}) denotes the active PUs' faded signal arriving at $\sr_{j}$, and we show the average power level of $s_j$ with $\sigma_{s,j}^2$.

We further consider a block-fading channel, and assume that the fading coefficients stay constant for a frame duration of $T$ seconds and change independently from one frame to another. In addition, we also assume that the activities of the PUs stay the same in each frame and change likewise independently from one frame to another. We further emphasize that the probability of the PUs being active in one frame is denoted by $\rho$. At the same time, the SRs experience the interference caused by the PUs contemporaneously when the PUs are active. We finally underline that the available bandwidth is $B$ Hz, so is the symbol rate assumed to be $B$ complex symbols per second.

\section{Channel Sensing and Power Constraints}\label{channelsensing_power}
\subsection{Channel Sensing}
The SUs (i.e., $\text{SU}_{1}$: $\sr_{1}$, $\text{SU}_{2}$: $\sr_{2}$, and $\text{SU}_{3}$: ST) operate channel sensing collaboratively before data transmission in order to detect the PUs. In more details, each SU initially performs channel sensing and obtains a sensing decision individually, and then, the SUs gather these channel sensing decisions at ST where the final channel sensing decision is determined\footnote{We assume that the channel sensing results are fed to ST over delay-and-error-free channels.}. Considering that the transmission strategies of the PUs are not known, an energy-based detection is applied at each SU. Therefore, the first $N$ seconds of the frame duration $T$ seconds are allocated for channel sensing. Noting that there are $\nu = N \times B$ complex symbols in a duration of $N$ seconds, the hypothesis testing problem between the noise $n_{l}(k)$ and the received signal $s_{l}(k)$ at $\text{SU}_{l}$ can be mathematically expressed as follows:
\begin{equation}\label{Hypothesis Test}
\begin{aligned}
\mathcal{H}_i &: y_l(k) = n_l(k), \quad k = 1,2, ... ,\nu,  \\
\mathcal{H}_b &: y_l(k) = n_l(k) + s_l(k), \quad k = 1,2, ... ,\nu,
\end{aligned}
\end{equation}
where $\mathcal{H}_{i}$ and $\mathcal{H}_{b}$ denote the true hypothesis corresponding to idle and busy states, respectively\footnote{We define that the PUs are active in the busy state, whereas there is not any active PU in the idle state.}. Above, $y_l$ is the received signal at $\text{SU}_{l}$ where $l\in\{1,2,3\}$. Considering the above detection problem, the optimal Neyman-Pearson detector at each SU is given by \cite{poor_book}
\begin{equation}
T_l(y) = \frac{1}{\nu} \sum_{n=1}^{\nu} |y_l(n)|^2 \overset{\hat{\mathcal{H}_b}}{\underset{\hat{\mathcal{H}_i}}{\gtrless}} \lambda_{l}
\end{equation}
where $\lambda_{l}$ is the detection threshold at each SU. Assuming that $\nu$ is sufficiently large, we can approximate $T_l(y)$ as a Gaussian random variable by invoking Central Limit Theorem. Now, it can be easily confirmed that $\mathbb{E}\{T_{l}(y)\}=\sigma_{n,l}^{2}$ and $\mathbb{E}\{T_{l}(y)\}=\sigma_{n,l}^{2}+\sigma_{s,l}^{2}$ when channel state is $\mathcal{H}_i$ and $\mathcal{H}_b$, respectively \cite{liang}. With these characterizations and Gaussian assumptions, we have the following probabilities of false alarm and detection at each SU in terms of $Q$-functions \cite{rahul}:
\begin{align*}
P_{f}^{l}&=\mathit{Q}\left(\frac{\lambda_{l}-\sigma_{n,l}^2}{\sqrt{\frac{2}{\nu}} \sigma_{n,l}^2} \right)\text{ and }P_{d}^{l}&=\mathit{Q}\left(\frac{\lambda_{l}-\sigma_{n,l}^2-\sigma_{s,l}^2}{\sqrt{\frac{2}{\nu}}(\sigma_{n,l}^2+\sigma_{s,l}^2)}\right)
\end{align*}
where $Q(x)=\frac{1}{\sqrt{2\pi}}\int_{x}^{\infty}e^{-x^{2}/2}dx$.

Moreover, we consider a hard-decision combining algorithm applied at ST, in which decision of each SU (either 0 or 1) is combined for the final channel sensing decision\footnote{0 indicates that the channel is idle ($\mathcal{H}_{i}$), whereas 1 indicates that the channel is busy ($\mathcal{H}_{b}$).}. Assuming that the SUs receive the signal emitted by the PUs at the same average power level (i.e., $\sigma_{s,l}^{2}=\sigma_{s}^{2}$), and that they have the same average noise variance (i.e., $\sigma_{n,l}^{2}=\sigma_{n}^{2}$), we consider the same detection threshold applied at each SU where $\lambda_{l}=\lambda$. As a result, we will have the same values for the probabilities of false alarm and detection at each SU: $P_{f}^{l}=P_{f}$ and $P_{d}^{l}=P_{d}$. In addition, we consider three different hard-decision algorithms at ST such as OR, Majority, and AND rules: Channel is considered as busy, when at least one SU detects it as busy with OR rule, or when at least two SUs detect it as busy with Majority rule, or when all of the SUs detect it as busy with AND rule. Given the above conditions, we can express the final probabilities of false-alarm and detection for each hard-decision algorithm as follows:
\begingroup
\allowdisplaybreaks
\begin{align}\label{Pf_Pd}
P_f^{final}&=\sum_{i=K}^{3}\dbinom{3}{i}(P_{f})^i (1-P_{f})^{3-i}, \\
P_d^{final}&=\sum_{i=K}^{3}\dbinom{3}{i}(P_{d})^i (1-P_{d})^{3-i}.
\end{align} 
\endgroup
where $K=1$, $K=2$, and $K=3$, when OR, Majority, and AND rules are applied, respectively.

\subsection{Interference Power Constraints}
Recall that ST chooses the transmission power policies with respect to the channel sensing results. In more details, if the channel is sensed as busy, ST sends the data symbol $x$ with the instantaneous transmission power policy $P^{b}(\mathbf{z})$, and $P^{b}(\mathbf{z})=P_1^b(\mathbf{z})+P_2^b(\mathbf{z})$ where $P_{1}^{b}(\mathbf{z})$ and $P_{2}^{b}(\mathbf{z})$ are the instantaneous power allocation policies for the users $\sr_{1}$ and $\sr_{2}$, respectively. On the other hand, when the channel is sensed as idle, the instantaneous transmission power policy is $P^{i}(\mathbf{z})$, and $P^{i}(\mathbf{z})=P_1^i(\mathbf{z})+P_2^i(\mathbf{z})$ where $P_{1}^{i}(\mathbf{z})$ and $P_{2}^{i}(\mathbf{z})$ are the instantaneous power allocation policies for $\sr_{1}$ and $\sr_{2}$, respectively. Note that $\mathbf{z}=\{z_1,z_2\}$ is the channel state vector. As a result of the channel sensing errors, we notice that ST deploys both policies $P^b(\mathbf{z})$ and $P^i(\mathbf{z})$ when the PUs are actually active. In particular, the policy $P^b(\mathbf{z})$ is deployed by ST with probability $P_d$, while the policy $P^i(\mathbf{z})$ is deployed with probability $(1-P_d)$ during the activities of the PUs. Therefore, in order to limit the interference caused by ST on the PUs, we impose the following combined interference power constraint on the SUs:
\begin{equation}\label{power_combined}
P_d\mathbb{E}_{\mathbf{z}}\{P^{b}(\mathbf{z})\}+(1-P_d)\mathbb{E}_{\mathbf{z}}\{P^i(\mathbf{z})\}\leq P_{int}
\end{equation}
where $P_{int}$ is the average interference power constraint\footnote{$P_{int}$ is the average interference power normalized over average fading power and path loss of the channels between ST and the primary receivers.}$^{,}$\footnote{A better interference protection strategy for the PUs can be realized by applying a peak power constraint on ST.}. In the sequel, we will be considering the following normalized instantaneous transmission power policies: $\mu^b(\mathbf{z})=\frac{P^b(\mathbf{z})}{P_{int}}$, $\mu^i(\mathbf{z})=\frac{P^i(\mathbf{z})}{P_{int}}$, $\mu_j^b(\mathbf{z})=\frac{P_j^i(\mathbf{z})}{P_{int}}$ and $\mu_j^i(\mathbf{z})=\frac{P_j^i(\mathbf{z})}{P_{int}}$. Note that $\mu^b(\mathbf{z})=\mu_1^b(\mathbf{z})+\mu_2^b(\mathbf{z})$ and $\mu^i(\mathbf{z})=\mu_1^i(\mathbf{z})+\mu_2^i(\mathbf{z})$. Finally, since the transmission power of ST is limited by $P_{int}$, we define the signal-to-noise ratio as $\tSNR=\frac{P_{int}}{B \sigma_n^2}$.

\section{Instantaneous Transmission Rates}\label{Rates}
Regarding the channel sensing result and its correctness, we have four different transmission scenarios:
\begin{enumerate}
\item Channel is busy, sensed as busy (correct detection),
\item Channel is busy, sensed as idle (miss-detection),
\item Channel is idle, sensed as busy (false alarm),
\item Channel is idle, sensed as idle (correct detection).
\end{enumerate}
We can easily see that ST will send with the power policy $\mu_j^b(\mathbf{z})$ for $\sr_{j}$ in Scenarios 1 and 3, and $\mu_j^i(\mathbf{z})$ in Scenarios 2 and 4. Therefore, assuming the interference caused by the primary users as additional Gaussian noise, the instantaneous ergodic channel capacity at each SR (i.e., $\sr_{j}$) during one transmission frame in each scenario can be expressed as follows \cite{goldsmith}:
\begin{equation}
\label{inst capacities}
C_{j,\tau}(\mathbf{z})=B\log_2\{1+\zeta_{j,\tau}(\mathbf{z})\}\text{ bits/sec for }\tau={1,2,3,4,} 
\end{equation}
where
\begin{align*}
& \zeta_{j,1}(\mathbf{z})=\frac{\mu_j^b(\mathbf{z})\tSNR z_j}{\beta+\tSNR\mu_m^b(\mathbf{z})z_j\mathbf{1}[z_m > z_j]},\\
& \zeta_{j,2}(\mathbf{z})=\frac{\mu_j^i(\mathbf{z})\tSNR z_j}{\beta+\tSNR\mu_m^i(\mathbf{z})z_j\mathbf{1}[z_m > z_j]},\\
& \zeta_{j,3}(\mathbf{z})=\frac{\mu_j^b(\mathbf{z})\tSNR z_j}{1+\tSNR\mu_m^b(\mathbf{z})z_j\mathbf{1}[z_m > z_j]},\\
& \zeta_{j,4}(\mathbf{z})=\frac{\mu_j^i(\mathbf{z})\tSNR z_j}{1+\tSNR\mu_m^i(\mathbf{z})z_j\mathbf{1}[z_m > z_j]},\\
\end{align*}
and $m\neq j$ for $m,j\in\{1,2\}$. Above, $\beta = 1 + \frac{\sigma_s^2}{\sigma_n^2}$, and $\mathbf{1}[\cdot]$ is an indicator function where $\mathbf{1}[a]=1$ if $a$ is true, and 0 otherwise. Note that we acquire an i.i.d. Gaussian codebook for the input symbols to the channel.

Since the SUs rely on channel sensing results with errors, they can not determine which scenario they are in. However, they know that they are in either Scenario 1 or 3, if the channel is sensed as busy, and that they are in either Scenario 2 or 4, if the channel is sensed as idle. Now, considering the above conditions, and assuming that ST performs linear (superposition) coding after obtaining channel side information, and that $\sr_{1}$ and $\sr_{2}$ apply successive decoding, one of the best transmission strategies could be that ST sends data at rates equal to $R_{j}^{b}(\mathbf{z})=C_{j,1}(\mathbf{z})$ when the channel is sensed as busy, i.e.,
\begin{equation}\label{rates_busy}
R_j^b(\mathbf{z})=B\log_2\left(1+\zeta_{j,1}(\mathbf{z})\right)
\end{equation}
for each $\sr_{j}$. We can easily observe that in Scenario 1, the instantaneous transmission rates are equal to the instantaneous channel capacities for both $\sr_{1}$ and $\sr_{2}$. Hence, there will be a reliable transmission to both receivers. Similarly, since the instantaneous transmission rates are less than or equal to the instantaneous channel capacities for both receivers, i.e., $R_{j}^{b}(\mathbf{z})\leq C_{j,3}(\mathbf{z})$, data will be reliably transmitted to the receivers in Scenario 3 as well. As a result, in one transmission frame, there will be $TR_{j}^{b}(\mathbf{z})$ bits transmitted effectively to the receivers when the channel is sensed as busy.

On the other hand, when the channel is sensed as idle, we consider the following two possible instantaneous transmission rate strategies:
\subsubsection{Strategy 1}
ST sends data with rates equal to $R_{j}^{i}(\mathbf{z})=C_{j,4}(\mathbf{z})$, i.e., 
\begin{equation}\label{rates_idle_1}
R_j^i(\mathbf{z})=B\log_2\left(1+\zeta_{j,4}(\mathbf{z})\right).
\end{equation}
This can be considered as a greedy transmission rate strategy, since ST sends data to each receiver at a rate equal to $\max\{C_{j,2}(\mathbf{z}),C_{j,4}(\mathbf{z})\}$. Moreover, we can easily observe that we always have $C_{j,2}(\mathbf{z})\leq C_{j,4}(\mathbf{z})$. However, this transmission strategy has a risk of transmission outage in Scenario 2 since $R_j^i(\mathbf{z})=C_{j,4}(\mathbf{z})\geq C_{j,2}(\mathbf{z})$. As a result, we assume that there will be no reliable transmission to the receivers in Scenario 2\footnote{It is assumed that a simple automatic repeat mechanism is incorporated in order to ensure that the erroneous data is retransmitted.}. Hence, when ST transmits with this strategy, the effective rate is $TR_{j}^{i}(\mathbf{z})$ bits per frame in Scenario 4, while it is 0 in Scenario 2.
\subsubsection{Strategy 2}
ST sends data with rates equal to $R_{j}^{i}(\mathbf{z})=C_{j,2}(\mathbf{z})$, i.e.,
\begin{equation}\label{rates_idle_2}
R_j^i(\mathbf{z})=B\log_2\left(1+\zeta_{j,2}(\mathbf{z})\right).
\end{equation} 
This strategy can be considered as a precautious transmission rate strategy, since ST sends data at a rate equal to $\min\{C_{j,2}(\mathbf{z}),C_{j,4}(\mathbf{z})\}$, which is always $C_{j,2}(\mathbf{z})$. Since ST sends data with lower rates in contrast to the rates in \emph{Strategy 1}, there will be no outage, and reliable transmission will be provided to SRs in both Scenarios 2 and 4. Therefore, the effective rate is $TR_{j}^{i}(\mathbf{z})$ bits per frame in Scenarios 2 and 4.

\section{Effective Capacity}\label{effect}
Effective capacity was defined by Wu \emph{et~al.} as the maximum constant arrival rate that a given service process can support while satisfying statistical QoS constraints specified by the QoS exponent $\theta$ \cite{Wu2003}. If we denote the stationary queue length by $Q$, then the decay rate of the tail distribution of the queue length $Q$ is defined by $\theta$:
\begin{equation}
\lim_{q \to \infty} \frac{\log\Pr(Q \geq q)}{q}=-\theta.
\end{equation}
Thus, we have the following approximation for the buffer violation probability for large $q_{max}:\Pr(Q\geq q_{max})\approx e^{-\theta q_{max}}$. Therefore, larger $\theta$ corresponds to more strict QoS constraints, while smaller $\theta$ implies looser constraints. Hence, effective capacity can provide us the maximum arrival rate to a data buffer when the system is subject to the statistical buffer constraints. And, for a given QoS exponent $\theta$, effective capacity is given by
\begin{equation}
C_{E}(\theta)=-\lim_{t\to\infty}\frac{1}{\theta t}\log_e\mathbb{E}\{e^{-\theta S(t)}\}
\end{equation}
where $S(t) = \sum_{l=1}^t r(l)$ is the time-accumulated service process, and $r(l)$ for $l=1,2,...$ is the discrete-time, stationary and ergodic stochastic service process. 

Noting that ST has two different transmission queues for storing the data allocated for each receiver $\sr_{1}$ and $\sr_{2}$, we consider that each queue has its own QoS constraints. Therefore, we denote the QoS exponent for each queue by $\theta_{j}$. Following the definition provided in \cite{Deli2013}, we can express the following effective capacity region for the above cognitive radio broadcast channel as follows:
\begin{align}\label{effect_region}
\mathcal{C}_E (\Theta) = \bigcup_{R_1,R_2} & \left \{ {C(\Theta)} 	\geq {\bf 0}: {C_j(\theta_j)} \right. \notag \\
& \left.\vphantom \leq \leq - \frac{1}{\theta_j T B} \log_e \mathbb{E_{\bf z}} \left \{ e^ {-\theta_j (T-N) R_j} \right \}\right  \}
\end{align}
where $\Theta = (\theta_1,\theta_2)$. Recall that the channel fading coefficients change from one frame to another independently, and similarly the activities of the PUs in one transmission frame do not depend on their activities in the previous frames. Now, following the steps in \cite{Sami2010}, we can express the normalized effective capacity in bits/sec/Hz for each SR as follows:
\begin{align}\label{effect_1}
C_j(\theta_j)=\frac{\log_e\mathbb{E}\{p_{b}e^{-\theta_j(T-N)R_j^b}+p_4e^{-\theta_j(T-N)R_j^i}+p_2\}}{-\theta_jTB}
\end{align}
if ST employs \emph{Strategy 1} when channel is sensed as idle, and
\begin{align}\label{effect_2}
C_j(\theta_j)=\frac{\log_e\mathbb{E}\{p_be^{-\theta_j(T-N)R_j^b}+(p_2+p_4)e^{-\theta_j(T-N)R_j^i}\}}{-\theta_jTB} 
\end{align}
if ST employs \emph{Strategy 2} when channel is sensed as idle, where $p_b=\rho P_d+(1-\rho)P_f$, $p_2=\rho(1-P_d)$, and $p_{4}=(1-\rho)(1-P_f)$. In the rest of the paper, we will omit the function input $\textbf{z}$, unless it is necessary.

\section{Optimal Transmission Power Policies}\label{power}
After characterizing the effective capacity region, we turn our attention to the optimal transmission power policies that will maximize the expressions in (\ref{effect_1}) and (\ref{effect_2}). Since ST has to send data considering two different QoS exponents $\theta_j$, we assume that ST prioritize each user over the transmission power allocation policies. Hence, we reconsider the interference power constraint given in (\ref{power_combined}) as follows:
\begin{equation}\label{power_combined_updated}
P_d\mathbb{E}\{P^{b}_{j}\}+(1-P_d)\mathbb{E}\{P^i_{j}\}\leq\eta_jP_{int}
\end{equation}
where $\eta_1=\delta$, $\eta_2=1-\delta$, and $\delta\in[0,1]$. We can clearly see that ST divides the available average transmission power between $\sr_1$ and $\sr_2$ with a ratio defined as $\delta$. Now, normalizing (\ref{power_combined_updated}) over $P_{int}$, we obtain\footnote{We assume that depending on the channel conditions and QoS constraints of each receiver, ST can select the value of $\delta$.}
\begin{equation}\label{power_combined_updated_normalized}
P_d\mathbb{E}\{\mu^{b}_{j}\}+(1-P_d)\mathbb{E}\{\mu^i_{j}\}\leq\eta_j.
\end{equation}

\begin{theo}\label{optimal_power_allocation_policies}
The optimal power allocation policies, $\mu^b_j$ and $\mu^i_j$, that maximize the effective capacities given in (\ref{effect_2}) with respect to given constraints in (\ref{power_combined_updated_normalized}) are given by
\begin{align}\label{mu_j_b}
&\mu_{j}^{b}=\notag\\
&\begin{cases}
\frac{\beta}{\tSNR z_{j}}\left[\left(\frac{p_{b}z_{j}}{\gamma_{j}P_{d}\beta}\right)^{\frac{1}{\kappa_{j}+1}}-1\right]^{+},& z_{j}\geq z_{m},\\
\frac{\beta+\mu_{m}^{b}\tSNR z_{j}}{\tSNR z_{j}}\left[\left(\frac{p_{b}z_{j}}{\gamma_{j}P_{d}(\beta+\mu_{m}^{b}\tSNR z_{j})}\right)^{\frac{1}{\kappa_{j}+1}}-1\right]^{+}, &\text{otherwise,}
\end{cases}
\end{align}
when the channel is sensed as busy, and
\begin{align}\label{mu_j_i}
&\mu_{j}^{i}=\notag\\
&\begin{cases}
\frac{\beta}{\tSNR z_{j}}\left[\left(\frac{p_{i}z_{j}}{\gamma_{j}P_{m}\beta}\right)^{\frac{1}{\kappa_{j}+1}}-1\right]^{+},& z_{j}\geq z_{m},\\
\frac{\beta+\mu_{m}^{i}\tSNR z_{j}}{\tSNR z_{j}}\left[\left(\frac{p_{i}z_{j}}{\gamma_{j}P_{m}(\beta+\mu_{m}^{i}\tSNR z_{j})}\right)^{\frac{1}{\kappa_{j}+1}}-1\right]^{+}, &\text{otherwise,}
\end{cases}
\end{align}
when the channel is sensed as idle, where $[\cdot]^+=\text{max}(\cdot,0)$ is the maximum operator, and $m\neq j$ for $m,j\in\{1,2\}$, and $P_{m}=1-P_{d}$. Note that $p_{i}=p_{2}+p_{4}$ in (\ref{mu_j_i}). $\gamma_{j}$ for $j\in\{1,2\}$ is the power threshold value in the power adaptation policies, and it can be obtained from the average interference power constraint (\ref{power_combined_updated_normalized}) through numerical techniques. Moreover, we can easily state that the optimal transmission power policies that maximize the effective capacities given in (\ref{effect_1}) with respect to given constraints in (\ref{power_combined_updated_normalized}) are $\mu_{j}^{b}$ in (\ref{mu_j_b}) when the channel is sensed as busy, and 
\begin{align}\label{mu_j_i_2}
&\mu_{j}^{i}=\notag\\
&\begin{cases}
\frac{1}{\tSNR z_{j}}\left[\left(\frac{p_{4}z_{j}}{\gamma_{j}P_{m}}\right)^{\frac{1}{\kappa_{j}+1}}-1\right]^{+},& z_{j}\geq z_{m},\\
\frac{1+\mu_{m}^{i}\tSNR z_{j}}{\tSNR z_{j}}\left[\left(\frac{p_{4}z_{j}}{\gamma_{j}P_{m}(1+\mu_{m}^{i}\tSNR z_{j})}\right)^{\frac{1}{\kappa_{j}+1}}-1\right]^{+}, &\text{otherwise,}
\end{cases}
\end{align}
when the channel is sensed as idle. Note that $\gamma_{j}$ obtained for (\ref{effect_1}) will be different than $\gamma_{j}$ obtained for (\ref{effect_2}).
\end{theo}

\emph{Proof:} We will first obtain the optimal transmission power policies that maximize the expression (\ref{effect_2}). Since logarithm is a monotonic function, we can attain the optimal power policies from the following minimization problem:
\begin{align}\label{optim_eq}
\min_{\substack{P_d\mathbb{E}\{\mu^{b}_{j}\}+(1-P_d)\mathbb{E}\{\mu^i_{j}\}\leq\eta_j}}p_be^{-\theta_j(T-N)R_j^b}+p_ie^{-\theta_j(T-N)R_j^i}
\end{align}
where $p_{i}=p_{2}+p_{4}$. Recall that $p_{b}$, $p_{2}$ and $p_{4}$ are defined at the end of Section \ref{effect}. Note further that $R_j^b=B\log_2\left(1+\zeta_{j,1}\right)$ and $R_j^i=B\log_2\left(1+\zeta_{j,2}\right)$ as given in (\ref{rates_busy}) and (\ref{rates_idle_2}), respectively. It is obvious that the expression in (\ref{optim_eq}) is strictly convex and the constraint (\ref{power_combined_updated_normalized}) is linear with respect to $\mu_{j}^{b}$ and $\mu_{j}^{i}$ \cite{Tang}. Then forming the Lagrange setting and taking the derivatives with respect to $\mu_{j}^{b}$ and $\mu_{j}^{i}$, we obtain
\begin{equation}\label{der_1}
\alpha_{j}P_{d}=\frac{\kappa_{j}p_{b}\tSNR z_{j}}{\beta}\left(1+\frac{\mu_{j}^{b}\tSNR z_{j}}{\beta}\right)^{-\kappa_{j}-1}
\end{equation}
and
\begin{equation}\label{der_2}
\alpha_{j}P_{m}=\frac{\kappa_{j}p_{i}\tSNR z_{j}}{\beta}\left(1+\frac{\mu_{j}^{i}\tSNR z_{j}}{\beta}\right)^{-\kappa_{j}-1},
\end{equation}
when $z_{j}\geq z_{m}$, and
\begin{equation}\label{der_3}
\alpha_{j}P_{d}=\frac{\kappa_{j}p_{b}\tSNR z_{j}}{\beta+\tSNR\mu_{m}^{b}z_{j}}\left(1+\frac{\mu_{j}^{b}\tSNR z_{j}}{\beta+\tSNR\mu_{m}^{b}z_{j}}\right)^{-\kappa_{j}-1}
\end{equation}
and
\begin{equation}\label{der_4}
\alpha_{j}P_{m}=\frac{\kappa_{j}p_{i}\tSNR z_{j}}{\beta+\tSNR\mu_{m}^{i}z_{j}}\left(1+\frac{\mu_{j}^{i}\tSNR z_{j}}{\beta+\tSNR\mu_{m}^{i}z_{j}}\right)^{-\kappa_{j}-1},
\end{equation}
when $z_{j}<z_{m}$, where $\alpha_{j}$ is the Lagrangian multiplier and $\kappa_{j}=\frac{\theta_{j}(T-N)B}{\log_{e}2}$. Defining $\gamma_{j}=\frac{\alpha_{j}}{\kappa_{j}\tSNR}$, and solving (\ref{der_1}) and (\ref{der_3}), we obtain (\ref{mu_j_b}), and solving (\ref{der_2}) and (\ref{der_4}), we have (\ref{mu_j_i}). Since we assume that all available transmission power should be used in general, we obtain the Lagrangian multiplier $\alpha_{j}$, and hence $\gamma_{j}$, numerically from the equality:
\begin{equation*}
P_d\mathbb{E}\{\mu^{b}_{j}\}+(1-P_d)\mathbb{E}\{\mu^i_{j}\}=\eta_j.
\end{equation*}

As for the transmission power policies that maximize the expression in (\ref{effect_1}), we similarly consider the following minimization problem:
\begin{align}\label{optim_eq_2}
\min_{\substack{P_d\mathbb{E}\{\mu^{b}_{j}\}+(1-P_d)\mathbb{E}\{\mu^i_{j}\}\leq\eta_j}}p_be^{-\theta_j(T-N)R_j^b}+p_4e^{-\theta_j(T-N)R_j^i}
\end{align}
where $R_j^b=B\log_2\left(1+\zeta_{j,1}\right)$ and $R_j^i=B\log_2\left(1+\zeta_{j,4}\right)$ as given in (\ref{rates_busy}) and (\ref{rates_idle_1}), respectively. Notice that since the transmission rate policies are different when the channel is sensed as idle while obtaining the effective capacity values in (\ref{effect_1}) and (\ref{effect_2}), the rates $R_j^i$ are different in the minimization problems (\ref{optim_eq}) and (\ref{optim_eq_2}). Again, after setting the Lagrangian function, when we take the derivatives with respect to $\mu_{j}^{b}$ and $\mu_{j}^{i}$, we will obtain the expressions in (\ref{der_1}) and (\ref{der_3}), and the expressions (\ref{der_2}) and (\ref{der_4}) with parameters $\beta$ and $p_{i}$ replaced with 1 and $p_{4}$, respectively. Solving these equalities, we will obtain the same formulation given in (\ref{mu_j_b}) and the one in (\ref{mu_j_i_2}).$\hfill{\square}$

\section{Numerical Results}\label{numeric}
\begin{figure}
\begin{center}
\includegraphics[width=\figsize\textwidth]{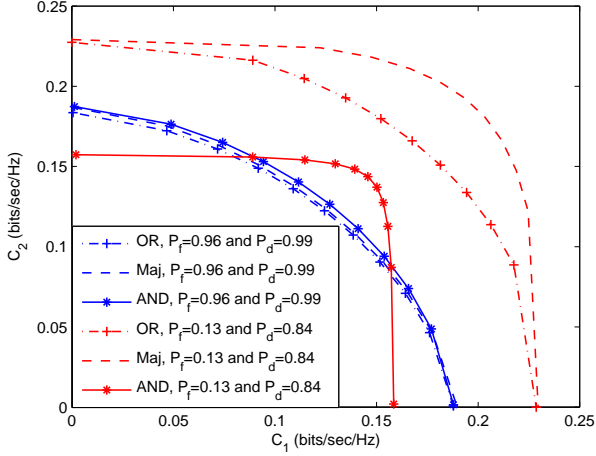}
\caption{Effective capacity region, $C_{1}$ vs. $C_{2}$, considering different hard-decision channel sensing algorithms when $(P_{f},P_{d})=(0.13,0.84)$ and $(P_{f},P_{d})=(0.96,0.99)$.}\label{Region_Combining}
\end{center}
\end{figure}
In this section, we provide the numerical results. Unless indicated otherwise, we consider the following parameter values. We consider a Rayleigh fading environment in which $z_1$ and $z_2$ are independent exponential random variables with $\mathbb{E} \{z_1\} = \mathbb{E} \{z_2\} = 1$. The available channel bandwidth is assumed to $B=2$ kHz, and the transmission frame duration is $T=1$ second while the duration $N=0.01$ seconds is allocated for channel sensing. We further assume that the PUs are active in one frame with probability $\rho=0.1$.

In Fig. \ref{Region_Combining}, we plot the effective capacity region considering different hard-decision algorithms with two different $(P_{f},P_{d})$ pairs at the SUs when both receivers have QoS exponents equal to $\theta_{1}=\theta_{2}=0.01$, and the interference-to-noise parameter is set to $\beta=2$, and the signal-to-noise ratio is $\tSNR=0$ dB. Here, we note that we obtain two different $(P_{f},P_{d})$ pairs by adjusting the channel detection threshold, $\lambda$. We can easily see that when the probability of false alarm, $P_{f}$, at each SU is high, the effective capacity regions obtained by applying different hard-decision algorithms at ST are very close to each other. The performance differences among hard-decision algorithms are negligible. On the other hand, when $P_{f}$ is very low at the SUs while having considerably good probability of detection values, $P_{d}$, the effective capacity regions show different behaviors. For instance, the SUs can obtain very high effective capacity values for both SRs by employing Majority rule when compared with other decision rules. Nevertheless, the same performance increase is not observed when AND rule is applied.

\begin{figure}
\begin{center}
\includegraphics[width=\figsize\textwidth]{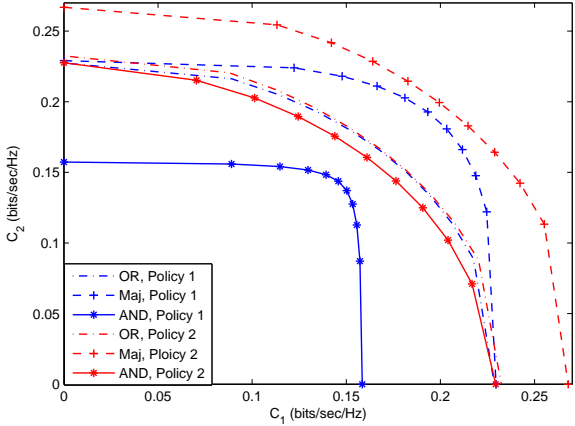}
\caption{Effective capacity region, $C_{1}$ vs. $C_{2}$, considering different instantaneous transmission rate policies, \emph{Strategy 1} and \emph{Strategy 2} when $(P_{f},P_{d})=(0.13,0.84)$.}\label{Region_options}
\end{center}
\end{figure}
\begin{figure}
\begin{center}
\includegraphics[width=\figsize\textwidth]{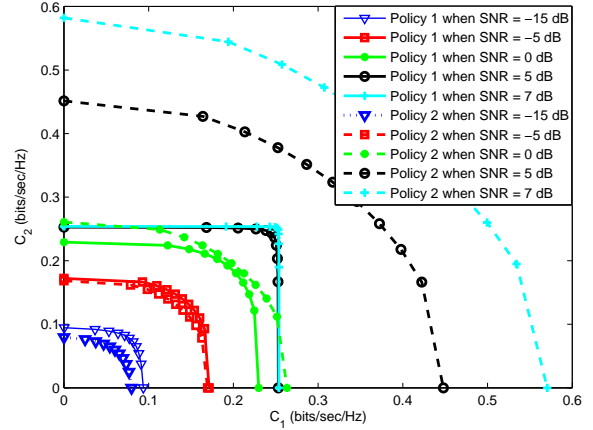}
\caption{Effective capacity region, $C_{1}$ vs. $C_{2}$, considering different instantaneous transmission rate policies, \emph{Strategy 1} and \emph{Strategy 2} when $(P_{f},P_{d})=(0.13,0.84)$ with different $\tSNR$ values by employing only Majority rule.}\label{Region_SNR}
\end{center}
\end{figure}
In Fig. \ref{Region_options}, we plot the effective capacity regions employing different instantaneous transmission rate policies when channel is sensed as idle with keeping the same channel parameters considered in Fig. \ref{Region_Combining}. We note that $(P_{f},P_{d})=(0.13,0.84)$. As seen in the figure, \emph{Strategy 2}, i.e., $R_{j}^{i}=C_{j,2}$, outperforms \emph{Strategy 1} in using any hard-decision algorithm when $\tSNR=0$ dB. Meanwhile, we can easily observe that the performance gap is very significant between \emph{Strategy 1} and \emph{Strategy 2} when Majority and AND rules are employed, while the performance is very low when OR rule is considered. We underline that ST does send data always with rates $R_{j}^{b}=C_{j,1}$ when the channel is detected as busy\footnote{We consider the different transmission rate policies that are employed only when the channel is sensed as idle.}. Hence, the effective capacity regions are calculated considering the expressions in (\ref{effect_1}) and (\ref{effect_2}) when \emph{Strategy 1} and \emph{Strategy 2} are employed, respectively, if the channel is sensed as idle. Furthermore, we show the effective capacity regions for different $\tSNR$ values (i.e., $\tSNR=7$, $5$, $0$, $-5$ and $-15$ dB) when only Majority rule is applied in Fig. \ref{Region_SNR}, in order to investigate the effects of employing \emph{Strategy 1} and \emph{Strategy 2} at loose and strict average interference power constraints. We can clearly see that when $\tSNR$ is high, \emph{Strategy 2} results in much higher performance levels when compared to \emph{Strategy 1}. In addition, the effective capacity region curve saturates after certain $\tSNR$ values when \emph{Strategy 1} is employed as seen when comparing the effective capacity values obtained at $\tSNR=5$ dB and $\tSNR=7$ dB, whereas the effective capacity values for both users increase with the increasing $\tSNR$ when \emph{Strategy 2} is employed. On the other hand, at low $\tSNR$ values, the performance of \emph{Strategy 1} is significantly higher than the performance of \emph{Strategy 2} as seen when we compare the outputs of both policies at $\tSNR=-5$ dB and $\tSNR=-15$ dB. We can conclude that the SUs should follow a greedy transmission rate strategy when $\tSNR$ is low, i.e., under strict average interference power constraints, while it is much more beneficial for the SUs to follow a transmission rate strategy that prevents data transmission outage when $\tSNR$ is high.

\begin{figure}
\begin{center}
\includegraphics[width=\figsize\textwidth]{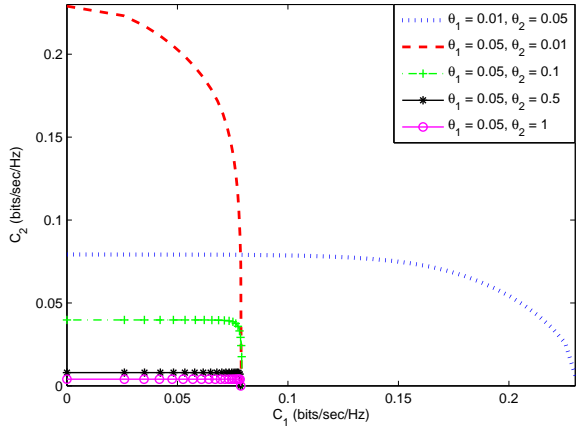}
\caption{Effective capacity region for different values of $\theta_1$ and $\theta_2$ when $\tSNR=0$ dB.}\label{Region_theta}
\end{center}
\end{figure}

\begin{figure}
\begin{center}
\includegraphics[width=\figsize\textwidth]{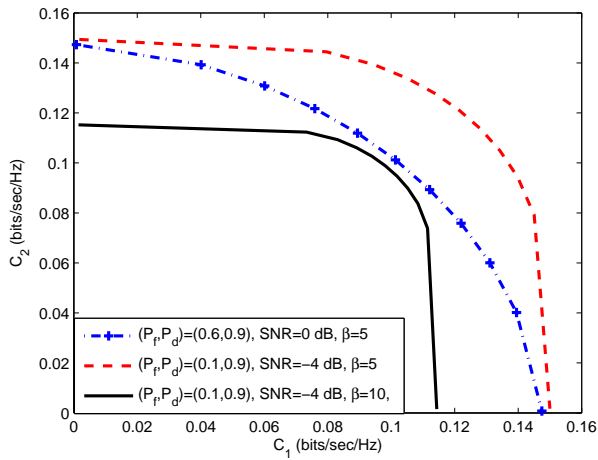}
\caption{Effective capacity region for different values of $\beta$ and ($P_{f}$,$P_{d}$) pairs.}\label{Region_beta}
\end{center}
\end{figure}

Moreover, we plot the effective capacity region regarding different QoS exponents for both users in Fig. \ref{Region_theta}. We can easily observe that with the increasing QoS constraints, there is a dramatic decline in the effective capacity region. However, we note that while keeping $\theta_1$ fixed and changing $\theta_2$, the maximum attainable effective capacity for the user $\sr_2$, which is obtained when the effective capacity of the user $\sr_1$ is going to zero, is not changing with different $\theta_{1}$ values. Finally, we plot the effective capacity region for different interference-to-noise parameter values, $\beta$ in Fig. \ref{Region_beta}, since $\beta$ is an important parameter that affects both the channel sensing performance and the effective capacity values. While the channel sensing performance is increasing with increasing $\beta$, the effective capacity is decreasing due to the decreasing transmission rates when the channel is sensed as busy, and the decreasing rates when the channel is sensed as idle assuming \emph{Strategy 2} is employed.

\section{Conclusion}\label{conclusion}
In this paper, we analyzed the effective capacities of cognitive radio broadcast channels under interference power constraints and channel uncertainty. Considering different cooperative channel sensing strategies, and different transmission rate selection strategies when the channel is sensed as idle, we formulated the effective capacity region of this broadcast channel model with two SRs and obtained the optimal transmission power policies that maximize this region. We showed that Majority rule outperforms the other sensing strategies in general, and that greedy and prudent transmission rate selection strategies are much more strategic when the interference power constraints are strict and loose, respectively in certain situations.

\end{document}